%% file: draft_BESIII_D0_K1ev.tex
\documentclass[aps,prl,twocolumn,showpacs,amsmath,amssymb]{revtex4-1}
\usepackage{amsmath}
\usepackage{graphicx}
\usepackage{subfigure}
\usepackage{epstopdf}
\usepackage{color}
\usepackage{multirow}
\usepackage{setspace}
\usepackage{overpic}
\usepackage{amssymb}
\usepackage[dvipdfmx, bookmarksnumbered, pdfstartview=FitH,colorlinks,urlcolor=blue, citecolor=blue,linkcolor=blue] {hyperref}
\usepackage{lineno}
\usepackage{bm}
\usepackage{rotating}
\usepackage[utf8]{inputenc}

\let\oldequation\equation
\let\oldendequation\endequation

\renewenvironment{equation}
  {\linenomathNonumbers\oldequation}
  {\oldendequation\endlinenomath}

\begin{document}

\title{\bf \boldmath
Observation of $D^0\to K_1(1270)^- e^+\nu_e$}

\input{BESIII_authors_2020_12_18}

\begin{abstract}
Using 2.93~fb$^{-1}$ of $e^+e^-$ collision data
taken with the BESIII detector at a center-of-mass energy of 3.773 $\rm \,GeV$,
the observation of the $D^0\to K_1(1270)^- e^+\nu_e$ semileptonic decay is presented.
The statistical significance of the decay $D^0\to K_1(1270)^- e^+\nu_e$ is greater than $10\sigma$.
The branching fraction of $D^0\to K_1(1270)^- e^+\nu_e$ is measured to be $(1.09\pm0.13^{+0.09}_{-0.16}
    \pm 0.12)\times10^{-3}$. Here, the first uncertainty is statistical, the second is systematic, and
    the third originates from the assumed branching fraction of $K_1(1270)^- \rightarrow K^-\pi^+\pi^-$.
    The fraction of longitudinal polarization in $D^0\to K_1(1270)^- e^+\nu_e$ is determined
    for the first time to be $0.50\pm0.19_{\rm stat}\pm0.08_{\rm syst}$.
\end{abstract}


\maketitle

\oddsidemargin  -0.2cm
\evensidemargin -0.2cm

Semileptonic (SL) $D$ decays
offer a good testbed to understand nonperturbative strong-interaction dynamics in weak decays~\cite{isgw,isgw2}.
Studies of the SL $D^{0(+)}$ decays into the strange axial-vector mesons $K_1(1270)$ or $K_1(1400)$
are especially appealing.
Reference~\cite{wangwei} points out that the combined measurements of $D^{0(+)}\to \bar K_{1}(1270) \ell^+ \nu_\ell$
and $B\to K_1(1270)\gamma$ provide a possible way to determine the
photon polarization in $b\to s\gamma$ transitions without considerable 
theoretical ambiguity.
Knowledge of the $b\to s\gamma$ photon polarization
plays a unique role in probing right-handed couplings in new physics~\cite{wangwei,Atwood,Becirevic}.
As LHCb reported a large up-down asymmetry
in $B^- \to K^-_{\rm res}(\to K^- \pi^+\pi^- )\gamma$ in the $K^-\pi^+\pi^-$ invariant mass bin of [1.1, 1.3]~GeV/$c^2$ which is dominated by $K_{1}(1270)^-$ contribution~\cite{lhcbk1g}.
Therefore, the decay $D^{0}\to K_{1}(1270)^-(\to K^-\pi^+\pi^-) \ell^+ \nu_\ell$
is particularly desired to quantify the hadronic effects of $K_{1}(1270)^-\to K^-\pi^+\pi^-$.
Throughout this Letter, charged conjugated modes are always implied.

To date, the $K_1(1270)$ and $K_1(1400)$ mesons have been extensively investigated in $\tau$, $D$, $B$, and charmonium decays~\cite{4,5,6,7,8,14,jpsi,chic0,1,2}.
In theory, the physical mass eigenstates of $K_1(1270)$ and $K_1(1400)$ are
decomposed as mixtures of the $^1\rm P_1$ and $^3\rm P_1$ states with a mixing angle $\theta_{K_1}$.
Various approaches were proposed to extract $\theta_{K_1}$, but with very different results~\cite{18,19,20,21,22,24,26,27}.
Experimental measurements of $D^{0(+)}\to \bar K_{1}(1270) e^+ \nu_e$ offer deeper insight into the mixing angle $\theta_{K_1}$,
which is essential for reliable calculations describing the $\tau$~\cite{18}, $B$~\cite{20,B-K1g},
and $D$~\cite{D-AP,D-AP-1} decays involving $K_1$, and for investigations in the field of hadron spectroscopy~\cite{Burakovsky}.

The branching fractions (BFs) of $D^{0(+)}\to \bar K_{1}(1270) e^+ \nu_e$ have been computed with different models: the Isgur-Scora-Grinstein-Wise (ISGW) quark model~\cite{isgw} and its update, ISGW2~\cite{isgw2},
three-point QCD sum rules (3PSR)~\cite{Khosravi},
covariant light-front quark model (CLFQM)~\cite{theory}, and the light-cone QCD sum rules (LCSR)~\cite{formfactor,Momeni}.
The predicted BFs, which are sensitive to $\theta_{K_1}$ and its sign, vary from $10^{-3}$ to $10^{-2}$~\cite{Khosravi,theory,formfactor}.
Measurements of these decay BFs and related longitudinal polarization are key to testing different theoretical calculations and
understanding the weak-decay mechanisms of $D$ mesons.
For example, assuming isospin symmetry, the ratio of the partial decay widths for the SL $D^{0(+)}$ decays, which are both mediated
via $c \to s e^+ \nu_e$, is expected to be unity~\cite{soni2}. Measuring the BFs thus allows a test of isospin invariance in $D^{0(+)}\to \bar K_1(1270) e^+\nu_e$.
Large $D^{0}\to K_{1}(1270)^- \ell^+ \nu_\ell$
samples also supply a clean environment, with no additional hadrons in
the final state, to accurately determine the mass and width of $K_1(1270)$, and to explore the relative strengths and
phases of $K_1(1270)^-$ decays into various final states that differ considerably with its neutral counterpart $\bar K_1(1270)^0$, which currently all suffer large uncertainties.

An observation of $D^+ \to \bar K_1(1270)^0 e^+\nu_e$ was previously reported by BESIII~\cite{bes3-k1ev}.
However, the only evidence for $D^0 \to K_1(1270)^- e^+\nu_e$ was reported by CLEO~\cite{cleores}.
This Letter presents an observation of $D^0 \to K_1(1270)^-e^+\nu_e$
by using 2.93~fb$^{-1}$ of $e^+e^-$ collision data~\cite{lum} recorded at a
center-of-mass energy $\sqrt s=3.773$~GeV with the BESIII detector~\cite{BESIII}.

Details about the design and performance of the BESIII
detector are given in Ref.~\cite{BESIII}.
Simulated samples produced with a {\sc geant4}-based~\cite{geant4} Monte Carlo (MC) package, which
includes the geometric description of the BESIII detector and the
detector response, are used to determine the detection efficiency
and to estimate the backgrounds. The simulation includes the beam-energy spread and initial-state radiation (ISR) in the $e^+e^-$
annihilations modeled with the generator {\sc kkmc}~\cite{ref:kkmc}.
The inclusive MC samples consist of the production of the $D\bar{D}$
pairs, the non-$D\bar{D}$ decays of the $\psi(3770)$, the ISR
production of the $J/\psi$ and $\psi(3686)$ states, and the
continuum processes incorporated in {\sc kkmc}~\cite{ref:kkmc}.
The known decay modes are modeled with {\sc
evtgen}~\cite{ref:evtgen} using BFs taken from the
Particle Data Group~\cite{pdg2020}, and the remaining unknown decays
from the charmonium states with {\sc
lundcharm}~\cite{ref:lundcharm}. Final-state radiation (FSR)
from charged final-state particles is incorporated with the {\sc
photos} package~\cite{photos}.
The $D^0\to K_1(1270)^- e^+\nu_e$ decay is simulated with
the ISGW2 model~\cite{isgw2} and the $K_1(1270)^-$ is allowed to decay into
all intermediate processes with final state of $K^-\pi^+\pi^-$.
The $K_1(1270)^-$ resonance shape is parameterized by a relativistic Breit-Wigner function with mass of $(1.253\pm0.007)$~GeV/$c^2$ and width of $(90\pm20)$~MeV~\cite{pdg2020}.
The BFs of $K_1(1270)$ subdecays measured by Belle~\cite{non-resonance} are input,
since they give better data/MC consistency than those reported in Ref.~\cite{pdg2020}.

At $\sqrt s=3.773$~GeV, $\bar D^{0}$ and $D^0$ mesons are produced in pairs.
The momenta of $\bar D^{0}$ and $D^0$ are equal and in opposite directions.
This advantage allows to study the $D$ decays with the 
double-tag (DT) technique first developed by Mark III~\cite{dtref}.
The $\bar D^0$ mesons are reconstructed by their hadronic decays to $K^+\pi^-$,
$K^+\pi^-\pi^0$, and $K^+\pi^-\pi^-\pi^+$.
These inclusively selected events are referred to as single-tag (ST) $\bar D^0$ mesons.
In the presence of the ST $\bar D^0$ mesons, candidates for $D^0\to K_1(1270)^{-} e^+\nu_e$
are selected to form DT events.
For a given tag mode, the BF of $D^0\to K_1(1270)^- e^+\nu_e$,
${\mathcal B}_{\rm SL}$,  is obtained by
\begin{equation}
\label{eq:bf}
{\mathcal B}_{\rm SL} = N_{\mathrm{DT}}/(N_{\mathrm{ST}}\cdot \varepsilon_{\rm SL}\cdot {\mathcal B}_{\rm sub}),
\end{equation}
where $N_{\rm ST}$ and $N_{\rm DT}$ are the ST and DT yields in data,
$\varepsilon_{\rm SL} = \varepsilon_{\rm DT}/\varepsilon_{\rm ST}$ is the efficiency of detecting the SL decay in the presence of the ST $\bar D^0$, and
${\mathcal B}_{\rm sub}$ is the BF of $K_1(1270)^-\to K^-\pi^+\pi^-$.
$\varepsilon_{\rm ST}$ and
$\varepsilon_{\rm DT}$ are the efficiencies of selecting the ST and DT candidates, respectively.

This analysis uses the same selection criteria of $K^\pm$, $\pi^\pm$, and $\pi^0$ as in Refs.~\cite{BESIII:2018sjg,BESIII:2018ccy,BESIII:2018qmf, BES3_FSR}.
The ST $\bar{D}^0$ mesons are identified by the energy difference
$\Delta E\equiv E_{\bar D^0}-E_{\rm beam}$ and the
beam-constrained mass $M_{\rm
  BC}\equiv\sqrt{E_{\mathrm{beam}}^{2}-|\vec{p}_{\bar{D}^0}|^{2}}$,
where $E_{\mathrm{beam}}$ is the beam energy, $E_{\bar D^0}$ and
$\vec{p}_{\bar{D}^0}$ are the total energy and momentum of the ST
$\bar{D}^0$ in the $e^+e^-$ rest frame. If there are multiple combinations in an
event, the combination with the smallest $|\Delta E|$ is chosen for each tag
mode.
Combinatorial backgrounds in the $M_{\rm BC}$ distributions are suppressed by requiring
$\Delta E$ within $(-29,\,27)$,  $(-69,\,38)$, and $(-31,\,28)$~MeV for $\bar D^0\to K^+\pi^-$,
$K^+\pi^-\pi^0$, and $K^+\pi^-\pi^-\pi^+$, respectively.

\begin{figure}[htp]
  \centering
  \includegraphics[width=1.0\linewidth]{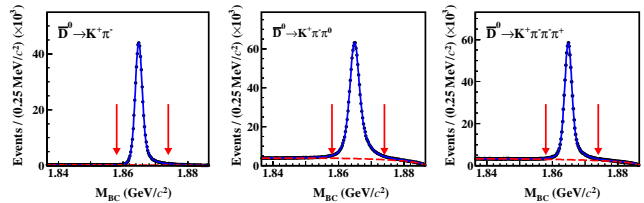}
  \caption{
Fits to the $M_{\rm BC}$ distributions of the ST candidates in data.
Points with error bars are data.
Blue solid curves are the fit results and
red dashed curves represent the background contributions of the fit.
Pair of red arrows in each subfigure indicate the $M_{\rm BC}$ window.
  }\label{fig:datafit_Massbc}
\end{figure}

The $M_{\rm BC}$ distributions of the accepted ST candidates in data for the three tag modes are shown in Fig.~\ref{fig:datafit_Massbc}.
To extract the ST yield for each tag mode, an unbinned maximum-likelihood fit is performed to the corresponding $M_{\rm BC}$ distribution.
The signal is described by the MC-simulated
shape convolved with a double-Gaussian function accounting for the
resolution difference between data and MC simulation, and the
background is modeled by an ARGUS function~\cite{ARGUS}.
Fit results are shown in Fig.~\ref{fig:datafit_Massbc}.
Events within $M_{\rm BC}\in (1.858,\,1.874)$ GeV/$c^2$ are kept for further analysis.
The ST yields for the $\bar D^0\to K^+\pi^-$, $K^+\pi^-\pi^0$, and $K^+\pi^-\pi^-\pi^+$ tag modes
are $542153\pm774_{\rm stat}$, $ 1080690\pm1727_{\rm stat}$, and $ 737036\pm1712_{\rm stat}$, respectively.

Particles recoiling against the ST $\bar D^0$ candidates are used
to reconstruct candidates for $D^0\to K_1(1270)^- e^+\nu_e$.
The $K_1(1270)^-$ is reconstructed via $K_1(1270)^-\to K^- \pi^+\pi^-$.
The $K^-$ and $\pi^\pm$ candidates are selected with the same criteria as the tag side.
Positron particle identification (PID) uses the combined information from the specific ionization energy loss ($dE/dx$), time of flight, and electromagnetic calorimeter (EMC),
with which we calculate the combined confidence levels under positron, pion, and kaon hypotheses $CL_e$, $CL_{\pi}$ and $CL_{K}$.
Positron candidate is required to satisfy $CL_e/(CL_e+CL_\pi+CL_K)>0.8$. To reduce backgrounds
from hadrons and muons, the positron candidate is required to satisfy
$E/p>0.8$, where $E$ is the energy deposited in the EMC and $p$ is the momentum measured by the multilayer drift chamber (MDC).
No additional charged track is allowed in the event.

To distinguish positrons from backgrounds related to hadrons, the positron candidates are required to satisfy
$E/p-0.38>0.14\times \chi^e_{dE/dx}$,
where $\chi^e_{dE/dx}$  is 
the $dE/dx$ $\chi^2$ with the positron hypothesis,
respectively.
To suppress the background from $D^0\to K^-\pi^+\pi^-\pi^+$,
we require $M_{K^-\pi^+\pi^-\pi^+_{e\to \pi}}<1.8~{\rm GeV}/c^2$,
where $\pi^+_{e\to \pi}$ is the positron candidate reconstructed with the pion mass hypothesis.
To suppress the background from $D^0\to K^-\pi^+\pi^0\,(\pi^0)$,
with $\pi^0\to e^+e^-\gamma$ (and missing another $\pi^0$), the opening angle between
$e^+$ and $\pi^-$ ($\theta_a$) is required to satisfy $\cos\theta_a < 0.94$.
To suppress the background from $D^0\to K^-\pi^+\pi^-\pi^+\pi^0$,
we require $M_{K^-\pi^+\pi^-\pi^+_{e\to \pi}\pi^0} < 1.4~{\rm GeV}/c^2$ when there is at least one reconstructed $\pi^0$ among the photons recoiling against the ST $\bar D^0$ meson in an event. Furthermore,
the opening angle between the missing momentum (defined below) and the most energetic unused shower ($\theta_b$) is required to satisfy
$\cos\theta_b<0.81$.
To suppress the background from $D^0\to K^-\pi^0e^+\nu_e$ with $\pi^0\to e^+e^-\gamma$,
we require $M_{\rm \pi^+\pi^-} > 0.31~{\rm GeV}/c^2$.  Background involving $K^0_S$ decay is suppressed by requiring $M_{\rm \pi^+\pi^-}$ outside the interval $(0.488, 0.508)$~{\rm GeV}/$c^2$.
For the $\bar{D}^0\to K^+\pi^-\pi^0$ tag mode, combinatorial background from $D^-\to K^+\pi^-\pi^-$ vs.~$D^+\to K^-\pi^+X$
is suppressed by requiring the difference between the beam-energy and the energy of the $(K^+\pi^-)_{\rm tag}\pi_{\rm sig}^-$
combination to be greater than 8~MeV.

Information concerning the undetectable neutrino is inferred by the kinematic quantity
$M^2_{\mathrm{miss}}\equiv E^2_{\mathrm{miss}}-|\vec{p}_{\mathrm{miss}}|^2$,
where $E_{\mathrm{miss}}$ and $\vec{p}_{\mathrm{miss}}$ are the missing energy and momentum
of the SL candidate, respectively,
calculated by $E_{\mathrm{miss}}\equiv E_{\mathrm{beam}}-\Sigma_j E_j$
and $\vec{p}_{\mathrm{miss}}\equiv-\vec{p}_{\bar D^0}-\Sigma_j \vec{p}_j$
in the $e^+e^-$ center-of-mass frame. The index $j$ sums over the $K^-$, $\pi^+$, $\pi^-$ and $e^+$ of the signal candidate,
and $E_j$ and $\vec{p}_j$ are the energy and momentum of the $j$-th particle, respectively.
To partially recover the energy lost to FSR and bremsstrahlung, the four-momenta of photon(s) within
$5^\circ$ of the initial positron direction are added to the positron four-momentum measured by the MDC.
To improve the $M^2_{\mathrm{miss}}$ resolution,
all the candidate tracks plus the missing neutrino are
subjected to a kinematic fit requiring energy
and momentum conservation, as well as the invariant
masses of the $\bar D^0$ and $D^0$ candidate particles being constrained to the nominal $D^0$ mass.
The momenta from the kinematic fit are used to calculate
$M^2_{\mathrm{miss}}$.

\begin{figure}[htbp]\centering
\includegraphics[width=1.0\linewidth]{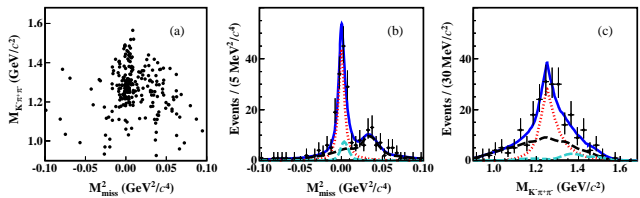}
\caption{
(a) Distribution of $M_{K^-\pi^+\pi^-}$ vs.~$M^2_{\rm miss}$ of the DT candidate events.
Projections of the 2D fit to (b) $M^2_{\rm miss}$ and (c) $M_{K^-\pi^+\pi^-}$.
The distributions are summed over all three tags.
In (b) and (c),
points with error bars are data;
blue solid, red dotted, green dashed, and black dashed curves are
total fit, signal, peaking background of $D^0\to K^-\pi^+\pi^+\pi^-$, and other background, respectively.
In (b), the peaking background concentrating around 0.033 GeV$^2/c^4$ is from $D^0\to K^-\pi^+\pi^+\pi^-\pi^0$.
}
\label{fig:fit}
\end{figure}

Figure~\ref{fig:fit}(a) shows the distribution of $M_{K^-\pi^+\pi^-}$ vs. $M^2_{\rm miss}$ of the accepted
$D^0\to K^-\pi^+\pi^-e^+\nu_e$ candidate events in data after combining all tag modes.
A clear signal, which concentrates around the $K_1(1270)^-$ nominal mass in the $M_{K^-\pi^+\pi^-}$ distribution and around zero in the
$M^2_{\rm miss}$ distribution, can be seen.
The DT yield is obtained from a two-dimensional (2D) unbinned extended maximum-likelihood simultaneous fit
to the data for the three tags.
In the fit,
the 2D signal shape is described by the MC-simulated shape extracted from the signal MC events of
$D^0\to K_1(1270)^- e^+\nu_e$.
The 2D shapes of the peaking background of $D^0\to K^-\pi^+\pi^+\pi^-$ and the other backgrounds are modeled by those derived from the inclusive MC sample.
The number of peaking background events from $D^0\to K^-\pi^+\pi^+\pi^-$ is fixed at the simulated value,
and the number of the other backgrounds is a free parameter.
The smooth 2D probability density functions of signal and background are modeled by using RooNDKeysPdf~\cite{roofit,class}.
The signal efficiencies with the ST modes $\bar D^0\to K^+\pi^-$, $K^+\pi^-\pi^0$, and $K^+\pi^-\pi^-\pi^+$
are $(14.08\pm0.14_{\rm stat})\%$, $(13.38\pm0.10_{\rm stat})\%$, and $(11.22\pm0.10_{\rm stat})\%$, respectively.
The BFs given by the three tags are constrained to have the same value in the fit.
The 2D fit projections to the $M^2_{\rm miss}$ and $M_{K^-\pi^+\pi^-}$ distributions
are shown in Figs.~\ref{fig:fit}(b) and \ref{fig:fit}(c), respectively.
From the fit, we obtain the DT yield of
$N_{\rm DT}= 109.0\pm12.5_{\rm stat}$.
The statistical significance of the signal is estimated to be greater than 10$\sigma$, by comparing the likelihoods with and
without the signal component, and taking the change in the number of degrees of freedom into account.
The fitted product of the BFs for $D^0\to K_1(1270)^- e^+\nu_e$ and $K_1(1270)^-\to K^-\pi^+\pi^-$ is
\begin{equation}
{\mathcal B}_{\rm SL}\cdot{\mathcal B}_{\rm sub} = (3.59\pm0.41_{\rm stat}{^{+0.31}_{-0.52}}_{\rm syst})\times10^{-4}.
\nonumber
\end{equation}
The reliability of the MC simulation is verified since
the data distributions of momenta and $\cos\theta$ of $K^-$, $\pi^+$, $\pi^-$ and $e^+$ as well as invariant masses of $K^-\pi^+$ and $\pi^+\pi^-$ are consistent with those of MC simulations.

In the BF measurement, the DT method ensures that most uncertainties arising from the ST selection cancel.
The uncertainty from the ST yield is assigned to be 0.5\%,
by examining the relative change in the yield between data and MC simulation after
varying the signal shape and the endpoint of the ARGUS function in the yield fits.

The systematic uncertainties originating from $e^+$ tracking and PID efficiencies
are studied by using the control samples of $e^+e^-\to\gamma e^+ e^-$ events and
those for $K^-$ and $\pi^\pm$ are investigated with the DT $D\bar D$ hadronic events.
All samples provide good coverage on track kinematics.
The $e^+$ efficiencies for tracking and PID are also re-weighted in 2D (momentum
and $\cos\theta$) to match those of the $D^0\to K_1(1270)^- e^+\nu_e$ data.
For $K^-$ and $\pi^+$, similar weighting is performed on momentum only
since the data and MC angular distributions already agree well.
Small differences between the data and MC efficiencies for $K^-$ tracking, $e^+$ tracking, and $e^+$ PID are found, which are
$+( 2.6\pm0.4)\%$, $+(1.0\pm0.2)\%$, and $-(1.4\pm0.2)\%$,
respectively.
The MC efficiencies, corrected by the aforementioned differences, are used for
the BF determination.
After corrections, the residual uncertainties related to the tracking (PID) efficiencies of $e^+$, $K^-$, $\pi^+$, and $\pi^-$
are assigned as 0.2\%\,(0.2\%), 0.4\%\,(0.3\%), 0.2\%\,(0.2\%), and 0.2\%\,(0.2\%), respectively.

Any systematic effects related to the requirements on
$M_{K^-\pi^+\pi^-\pi^+_{e\to \pi}}$,
$M_{K^-\pi^+\pi^-\pi^+_{e\to \pi}\pi^0}$,
$M_{\pi^+\pi^-}$,
$\Delta E[(K^-\pi^+)_{\rm tag}\pi_{\rm sig}^+]$,
$\cos\theta_a$, and
$\cos\theta_b$,
are examined by varying individual requirements by
$\pm0.05$ GeV/$c^2$,
$\pm0.05$ GeV/$c^2$,
$\pm0.01$ GeV/$c^2$,
$\pm0.004$ GeV,
$\pm0.02$, and $\pm0.02$,
respectively.
Accounting for correlations in the samples, the changes in the BFs are smaller than the statistical uncertainty on the difference,
so neither a systematic correction nor uncertainty is applied from this source according to Ref.~\cite{err}.
The systematic uncertainty from the input BFs of $K_1(1270)^-$ subdecays
is assigned to be 3.0\%
by varying each of the quoted subdecay BFs of Belle~\cite{non-resonance} by $\pm 1\sigma$ and
by comparing our nominal signal efficiency to the one based on the world average BFs of $K_1(1270)^-$ decays.

The systematic uncertainty of the 2D fit is estimated
to be $^{+6.9\%}_{-13.5\%}$ via two aspects.
The uncertainty from signal shape is mainly caused by varying the $K_1(1270)$ width by $\pm1\sigma$ ($\pm 6.0$\%).
The uncertainty of background shape is mainly due to non-$K_1(1270)^-$ sources
of $K^-\pi^+\pi^-$ ($^{+0.0\%}_{-8.7\%}$), which is the change of the fitted DT yield after fixing a non-resonant component by referring to the non-resonant fraction in $B \to J/\psi \bar K\pi\pi$~\cite{non-resonance}.
The uncertainty due to ignoring $D^+\to K_1(1400)^{-}e^+\nu_e$
is assigned as $^{+0.0\%}_{-7.6\%}$, by performing pseudoexperiments to evaluate fit biases and assuming its contribution is one order of magnitude lower than our signal decay~\cite{pdg2020,theory, formfactor}, 
while the effects from $D^0\to K^*(1410)^-e^+\nu_e$ and $D^0\to K_2^*(1430)^-e^+\nu_e$ are negligible.

The uncertainty due to the MC samples' limited size, 1.0\%, is considered as a source of systematic uncertainty.

The uncertainty from FSR recovery is assigned as 0.3\% by referring to Ref.~\cite{BES3_FSR}.
The uncertainty due to the kinematic fit is ignored since it is only used to improve the $M^2_{\rm miss}$ resolution.
The total systematic uncertainty is estimated to be $^{+8.7\%}_{-14.5\%}$ by
adding all the individual contributions in quadrature.

Using the world average of
${\mathcal B}_{\rm sub}= (32.9\pm3.6)\%$~\cite{pdg2020,K10},
we obtain
\begin{equation}
{\mathcal B}_{\rm SL} = {\mathcal B}_{D^0\to K_1(1270)^-e^+\nu_e} = (1.09\pm 0.13^{+0.09}_{-0.16} \pm 0.12)\times10^{-3},\nonumber
\end{equation}
where the third uncertainty is from the external uncertainty of the assumed BF ${\mathcal B}_{\rm sub}$.

\begin{figure}[htp]
  \centering
  \includegraphics[width=1.0\linewidth]{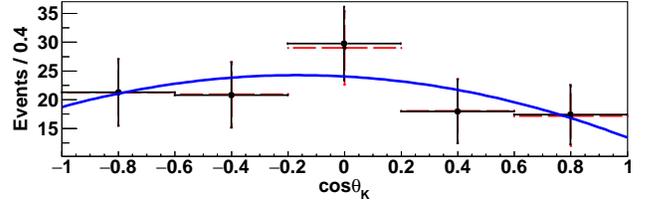}
  \caption{
      Fit to the efficiency corrected signal yields in bins of $\cos\theta_K$.
    Solid (dashed) lines with error bars are signal yields after (before)
    efficiency correction that accounts for efficiency differences between the first bin and other bins.
Blue solid curve is the fit result.
  }\label{fig:FLfit}
\end{figure}

A 2D fit is also performed in each of the five equal-sized $\cos\theta_K$ bins 
to determine the background subtracted angular distribution, where $\theta_K$ is the angle between the opposite of $D^0$ flight direction and the normal
$\vec{p}_{\pi,{\rm slow}}\times \vec{p}_{\pi,{\rm fast}}$ to the $K^-\pi^+\pi^-$ plane in the $K^-\pi^+\pi^-$ rest frame, where $\vec{p}_{\pi,{\rm slow}}$ ($\vec{p}_{\pi,{\rm fast}}$) is the momentum of the lower (higher) momentum pion~\cite{lhcbk1g, wangwei}.
Figure~\ref{fig:FLfit} shows the fit to the $\theta_K$ distribution with a second-order polynomial function~\cite{wangwei},
\begin{equation}
    \frac{d\Gamma\left(D^0\to K_1(1270)^-e^+\nu_e\right)}{d \cos\theta_K} \propto 1 + k_1 \cos\theta_K + k_2 \cos^2\theta_K, \label{eq:costhk}
\end{equation}
where $k_1$ is a free parameter, $k_2 = (1-3 F_L)/(1+ F_L)$, $F_L = \frac{|c_0|^2}{|c_0|^2+|c_+|^2+|c_-|^2}$ is the fraction of $K_1$ longitudinal polarization, with
$c_{0,\pm}$ representing the nonperturbative amplitudes for $D\to K_1$ with different polarizations.
As $\theta_K$ is parity odd, the sign for $\cos\theta_K$ in $\bar{D}^0$ decays is flipped.
We obtain $F_L = 0.50 \pm 0.17_{\rm stat} \pm 0.08_{\rm syst}$,
where the systematic uncertainty mainly comes from signal shape modeling.
Our $F_L$ result is compatible within 1$\sigma$ with the LCSR predictions 
in Ref.~\cite{formfactor}.

In summary, using 2.93~$\rm fb^{-1}$ of $e^+e^-$ collision data taken at $\sqrt s=3.773$~GeV,
we report the first observation of $D^0\to K_1(1270)^- e^+\nu_e$.
The obtained product of the BFs for $D^0\to K_1(1270)^- e^+\nu_e$ and $K_1(1270)^-\to K^-\pi^+\pi^-$ is consistent with the CLEO's result but with precision improved by about threefold~\cite{cleores}.
Our BF of $D^0\to K_1(1270)^- e^+\nu_e$ contributes $(1.68\pm0.35)\%$ of the total SL decay width of $D^0$~\cite{pdg2020}, which lies between the ISGW prediction (1\%) and the ISGW2 prediction (2\%), consistent with the BESIII result for the $D^+$ counterpart~~\cite{bes3-k1ev}.
Our BF of $D^0\to K_1(1270)^- e^+\nu_e$ agrees with
the CLFQM and LCSR predictions when $\theta_{K_1}\approx 33^\circ$ or $57^\circ$~\cite{theory,Momeni}
and clearly disfavors the prediction reported in Ref.~\cite{formfactor}.
Using the BF of $D^+\to \bar K_1(1270)^0e^+\nu_e$ measured by BESIII~\cite{bes3-k1ev} and the world-average lifetimes of $D^0$ and $D^+$~\cite{pdg2020}, we determine the ratio of the partial decay widths of the two decays to be $\Gamma_{D^0\to K_1(1270)^- e^+\nu_e}/\Gamma_{D^+\to \bar K_1(1270)^0 e^+\nu_e}=1.20\pm0.20\pm0.14\pm0.04$, where
the systematic uncertainties from the background shape, the tracking and PID efficiencies of $K^-$, $\pi^+$, and $e^+$ as well as FSR recovery  are canceled,
the uncertainties of the lifetimes of $D^0$ and $D^+$ are included; the uncertainties of the quoted BFs for $K_1(1270)$ decays are largely canceled.
This result agrees with unity as predicted by isospin symmetry. Our $F_L$ measurement is compatible with theoretical predictions. 

Further studies of the $K\pi\pi$ system with larger $D^{0(+)}\to \bar K_{1}(1270) e^+\nu_e$ samples
at BESIII in the near future~\cite{bes3-white-paper} and large 
$D^{*+}\to D^0 \pi^+,\ D^{0}\to K_{1}(1270)^-(\to K^-\pi^+\pi^-) \ell^+\nu_\ell$ samples 
at LHCb~\cite{adavis} allow to extract the hadronic-transition form factors, the mass, width, and 
the subdecay BFs of $K_1(1270)$, and to quantify the hadronic effects in $K_1(1270)\to K\pi\pi$.
These will benefit the precise determinations of photon polarization in $b\to s\gamma$ transitions
with high statistics sample of $B\to K_1(1270)\gamma$ at Belle~II~\cite{belle2-white-paper} and LHCb~\cite{lhcb-white-paper}, 
thereby effectively over-constraining the right-handed couplings in new physics models.

The BESIII collaboration thanks the staff of BEPCII and the IHEP computing center for their strong support. Authors thank Wei Wang and Fu-Sheng Yu for helpful discussions. This work is supported in part by National Key Research and Development Program of China under Contracts Nos. 2020YFA0406300, 2020YFA0406400; National Natural Science Foundation of China (NSFC) under Contracts Nos. 11775230, 11605124, 11625523, 11635010, 11735014, 11822506, 11835012, 11935015, 11935016, 11935018, 11961141012; the Chinese Academy of Sciences (CAS) Large-Scale Scientific Facility Program; Joint Large-Scale Scientific Facility Funds of the NSFC and CAS under Contracts Nos. U1732263, U1832207, U1932108; CAS Key Research Program of Frontier Sciences under Contracts Nos. QYZDJ-SSW-SLH003, QYZDJ-SSW-SLH040; 100 Talents Program of CAS; INPAC and Shanghai Key Laboratory for Particle Physics and Cosmology; ERC under Contract No. 758462; European Union Horizon 2020 research and innovation programme under Contract No. Marie Sklodowska-Curie grant agreement No 894790; German Research Foundation DFG under Contracts Nos. 443159800, Collaborative Research Center CRC 1044, FOR 2359, FOR 2359, GRK 214; Istituto Nazionale di Fisica Nucleare, Italy; Ministry of Development of Turkey under Contract No. DPT2006K-120470; National Science and Technology fund; Olle Engkvist Foundation under Contract No. 200-0605; STFC (United Kingdom); The Knut and Alice Wallenberg Foundation (Sweden) under Contract No. 2016.0157; The Royal Society, UK under Contracts Nos. DH140054, DH160214; The Swedish Research Council; U. S. Department of Energy under Contracts Nos. DE-FG02-05ER41374, DE-SC-0012069.

\end{document}

%% file: BESIII_authors_2020_12_18.tex
\author{
M.~Ablikim$^{1}$, M.~N.~Achasov$^{10,c}$, P.~Adlarson$^{67}$, S. ~Ahmed$^{15}$, M.~Albrecht$^{4}$, R.~Aliberti$^{28}$, A.~Amoroso$^{66A,66C}$, M.~R.~An$^{32}$, Q.~An$^{63,49}$, X.~H.~Bai$^{57}$
% ...
\\
\vspace{0.2cm}
(BESIII Collaboration)\\
\vspace{0.2cm} {\it
$^{1}$ Institute of High Energy Physics, Beijing 100049, People's Republic of China\\
% ...
$^{4}$ Bochum Ruhr-University, D-44780 Bochum, Germany\\
% ...
$^{10}$ G.I. Budker Institute of Nuclear Physics SB RAS (BINP), Novosibirsk 630090, Russia\\
% ...
$^{15}$ Helmholtz Institute Mainz, Johann-Joachim-Becher-Weg 45, D-55099 Mainz, Germany\\
% ...
$^{28}$ Johannes Gutenberg University of Mainz, Johann-Joachim-Becher-Weg 45, D-55099 Mainz, Germany\\
% ...
$^{32}$ Liaoning Normal University, Dalian 116029, People's Republic of China\\
% ...
$^{49}$ State Key Laboratory of Particle Detection and Electronics, Beijing 100049, Hefei 230026, People's Republic of China\\
% ...
$^{57}$ University of Jinan, Jinan 250022, People's Republic of China\\
% ...
$^{63}$ University of Science and Technology of China, Hefei 230026, People's Republic of China\\
% ...
$^{66}$ University of Turin and INFN, (A)University of Turin, I-10125, Turin, Italy; (B)University of Eastern Piedmont, I-15121, Alessandria, Italy; (C)INFN, I-10125, Turin, Italy\\
$^{67}$ Uppsala University, Box 516, SE-75120 Uppsala, Sweden\\
% ...
$^{c}$ Also at the Novosibirsk State University, Novosibirsk, 630090, Russia\\
% ...
}
}